\documentclass[aps,prl,twocolumn,superscriptaddress]{revtex4}
\usepackage{graphicx}
\usepackage{epstopdf}

\usepackage{amsmath} \usepackage{amssymb}

\begin{document}

\title{Electronic correlations and flattened band \\ in magnetic Weyl semimetal candidate Co$_{3}$Sn$_{2}$S$_{2}$}

\author{Yueshan Xu$^{\dagger}$}
\affiliation{Beijing National Laboratory for Condensed Matter Physics, Institute of Physics,
Chinese Academy of Sciences, Beijing 100190, China}
\affiliation{School of Physical Sciences, University of Chinese Academy of Sciences, Beijing 100190, China}

\author{Jianzhou Zhao$^{\dagger}$}
\affiliation{Co-Innovation Center for New Energetic Materials, Southwest University of Science and Technology, Mianyang, Sichuan 621010, China}
\affiliation{Physik-Institut, Universit$\ddot{a}$t Z$\ddot{u}$rich, Winterthurerstrasse 190, CH-8057 Zurich, Switzerland}

\author{Changjiang Yi$^{\dagger}$}
\affiliation{Beijing National Laboratory for Condensed Matter Physics, Institute of Physics,
Chinese Academy of Sciences, Beijing 100190, China}
\affiliation{School of Physical Sciences, University of Chinese Academy of Sciences, Beijing 100190, China}

\author{Qi Wang}
\affiliation{Beijing Key Laboratory of Opto-electronic Functional Materials and Micro-nano Devices, Department of Physics, Renmin University of China, Beijing 100872, China}

\author{Qiangwei Yin}
\affiliation{Beijing Key Laboratory of Opto-electronic Functional Materials and Micro-nano Devices, Department of Physics, Renmin University of China, Beijing 100872, China}

\author{Yilin Wang}
\affiliation{Department of Condensed Matter Physics and Materials Science, Brookhaven National Laboratory, Upton, New York 11973, USA}

\author{\\Xiaolei Hu}
\affiliation{Beijing National Laboratory for Condensed Matter Physics, Institute of Physics,
Chinese Academy of Sciences, Beijing 100190, China}
\affiliation{School of Physical Sciences, University of Chinese Academy of Sciences, Beijing 100190, China}

\author{Luyang Wang}
\affiliation{Sate Key Laboratory of Optoelectronic Materials and Technologies, School of Physics, Sun Yat-Sen University, Guangzhou 510275, China}

\author{Enke Liu}
\affiliation{Beijing National Laboratory for Condensed Matter Physics, Institute of Physics,
Chinese Academy of Sciences, Beijing 100190, China}
\affiliation{Songshan Lake Materials Laboratory, Dongguan, Guangdong 523808, China}

\author{Gang Xu}
\affiliation{Wuhan National High Magnetic Field Center,\\ Huazhong University of Science and Technology, Wuhan, Hubei 430074, China}

\author{Ling Lu}
\affiliation{Beijing National Laboratory for Condensed Matter Physics, Institute of Physics,
Chinese Academy of Sciences, Beijing 100190, China}
\affiliation{Songshan Lake Materials Laboratory, Dongguan, Guangdong 523808, China}

\author{Alexey A. Soluyanov}
\affiliation{Physik-Institut, Universit$\ddot{a}$t Z$\ddot{u}$rich, Winterthurerstrasse 190, CH-8057 Zurich, Switzerland}

\author{\\Hechang Lei}
\affiliation{Beijing Key Laboratory of Opto-electronic Functional Materials and Micro-nano Devices, Department of Physics, Renmin University of China, Beijing 100872, China}

\author{Youguo Shi}
\affiliation{Beijing National Laboratory for Condensed Matter Physics, Institute of Physics,
Chinese Academy of Sciences, Beijing 100190, China}
\affiliation{Songshan Lake Materials Laboratory, Dongguan, Guangdong 523808, China}

\author{Jianlin Luo}
\affiliation{Beijing National Laboratory for Condensed Matter Physics, Institute of Physics,
Chinese Academy of Sciences, Beijing 100190, China}
\affiliation{Songshan Lake Materials Laboratory, Dongguan, Guangdong 523808, China}

\author{Zhi-Guo Chen*}
\affiliation{Beijing National Laboratory for Condensed Matter Physics, Institute of Physics,
Chinese Academy of Sciences, Beijing 100190, China}
\affiliation{Songshan Lake Materials Laboratory, Dongguan, Guangdong 523808, China} 
 
\begin{abstract}  
\noindent \textbf{The interplay between electronic correlations and topological protection may offer a rich avenue for discovering emergent quantum phenomena in condensed matter. However, electronic correlations have so far been little investigated in Weyl semimetals (WSMs) by experiments. Here, we report a combined optical spectroscopy and theoretical calculation study on the strength and effect of electronic correlations in a magnet Co$_{3}$Sn$_{2}$S$_{2}$. The electronic kinetic energy estimated from our optical data is about half of that obtained from single-particle \textit{ab initio} calculations in the ferromagnetic ground state, which indicates intermediate-strength electronic correlations in this system. Furthermore, comparing the energy and side-slope ratios between the interband-transition peaks at high energies in the experimental and single-particle-calculation-derived optical conductivity spectra with the bandwidth-renormalization factors obtained by many-body calculations enables us to estimate the Coulomb-interaction strength (\textit{U} $\sim$ 4 eV) in Co$_{3}$Sn$_{2}$S$_{2}$. Besides, a sharp experimental optical conductivity peak at low energy, which is absent in the single-particle-calculation-derived spectrum but is consistent with the optical conductivity peaks obtained by many-body calculations with \textit{U} $\sim$ 4 eV, indicates that an electronic band connecting the two Weyl cones is flattened by electronic correlations and emerges near the Fermi energy in Co$_{3}$Sn$_{2}$S$_{2}$. Our work paves the way for exploring flat-band-generated quantum phenomena in WSMs.}   
\end{abstract} 
 

\maketitle
  
\begin{figure}
	\includegraphics[width=8.6cm]{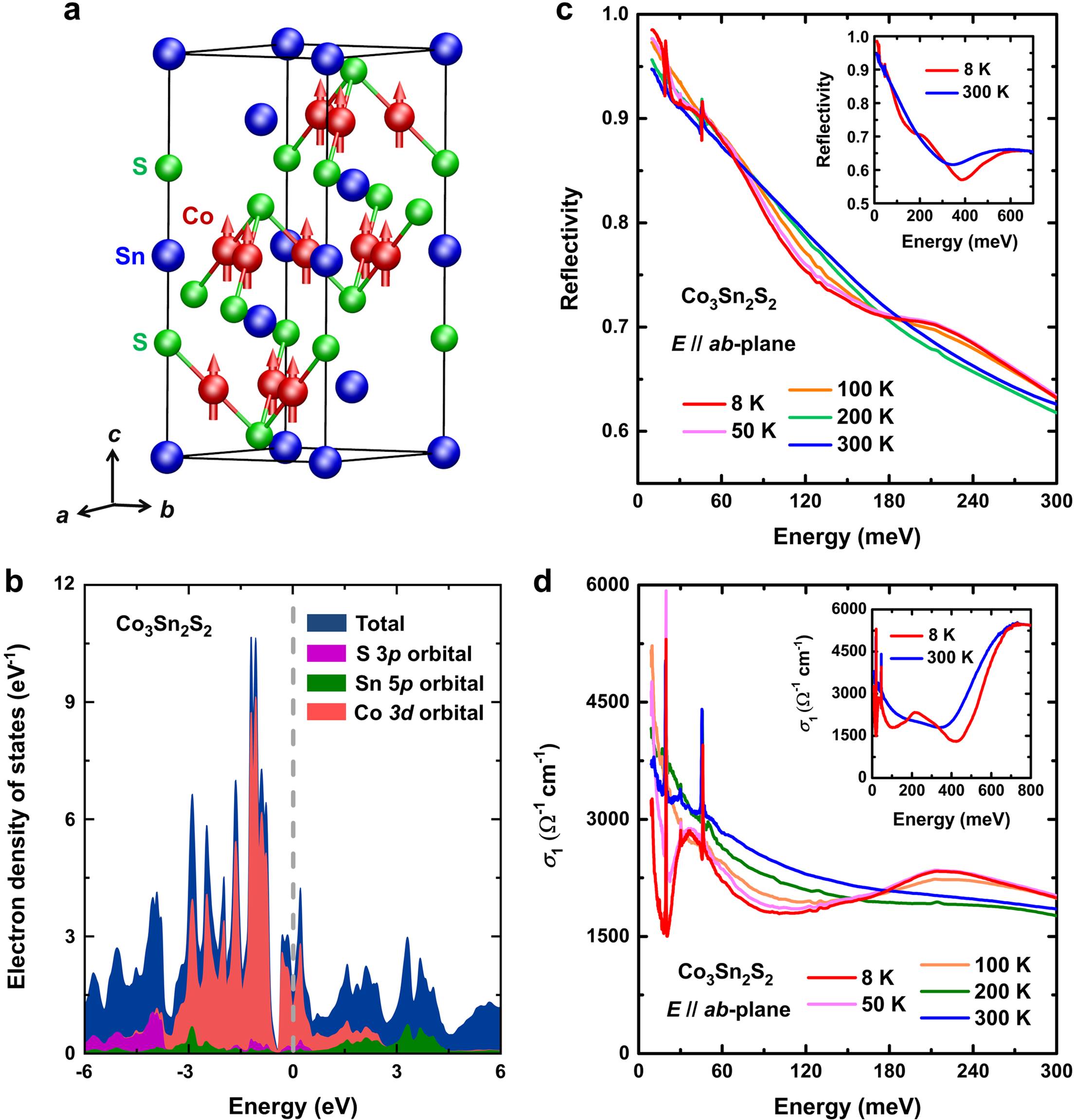}\\
	\centering
	\caption{\textbf{\textit{Ab}-plane optical response of Co$_{3}$Sn$_{2}$S$_{2}$.} \textbf{a}, Crystal structure of Co$_{3}$Sn$_{2}$S$_{2}$. A quasi-two-dimensional Co$_{3}$Sn layer is sandwiched between the sulfur atoms. The magnetic moments on the cobalt sites are along the \textit{c}-axis. \textbf{b}, Electron density of states (DOS) obtained by single-particle \textit{ab initio} calculations in the FM ground state. The electron DOS near the Fermi energy is mainly contributed by the electronic bands with Co 3\textit{d} orbital characters. \textbf{c}, Several representative reflectance spectra \textit{R}($\omega$) measured with the electric field (\textit{E}) of the incident light parallel to the crystalline \textit{ab}-plane. The inset shows that the \textit{R}($\omega$) up to 700 meV. \textbf{d}, Real parts $\sigma_{1}$(\emph{$\omega$}) of the \textit{ab}-plane optical conductivity at different temperatures. The inset displays the $\sigma_{1}$(\emph{$\omega$}) up to 800 meV.} 
	\label{FIG.1.linear}
\end{figure}  
   
\noindent Electronic correlations, which is a type of many-body interactions---Coulomb interactions between electrons, lie at the heart of condensed matter physics due to their crucial roles in producing a variety of novel quantum phenomena, such as unconventional superconductivity \cite{Bednorz, Hosono}, heavy-fermion behavior \cite{Stewart, Steglich}, and Mott insulation \cite{Imada, Kotliar, XGWen, Morosan}. Thus, theoretical predictions and experimental observations of topological quantum states in real materials with significant electronic correlations have generated tremendous interest in the scientific community \cite{SCZhang_TMI, Balents_NP, Coleman_TKI, Xiangang Wan}. Therein, Weyl semimetals (WSMs) represent a kind of topological quantum states which host pairs of bulk Weyl cones and surface Fermi arcs connecting pairs of Weyl points with opposite chirality \cite{Xiangang Wan, A. A. Burkov, Gang Xu, Hongming Weng, Shin-Ming Huang,Su-Yang Xu1, B. Q. Lv, Chandra Shekhar, L. X. Yang, Z. K. Liu, Soluyanov}. Recently, theoretical studies indicate that sufficiently strong electronic correlations can gap out bulk Weyl nodes and thus break WSM states \cite{Aji, Kim, ZWang_1, Abanin, Nomura, ZWang_2, HYao, Nagaosa, PYe, Rache, Roy}. Therefore, if a system which is predicted to exhibit a WSM phase in a non-interacting single-particle picture has nonnegligible electronic correlations, it will be significant to investigate the influence of electronic correlations on its predicted WSM state \cite{NPOng, Kuroda, DaiX, LaiHH, HQYuan}. Additionally, several correlated electron systems, such as kagome-lattice compounds \cite{Zengchanggan, JXYin} and heavy-fermion materials \cite{NLWang_1}, have been reported to host flat bands (i.e., dispersionless bands) which can provide a footstone for the emergence of various quantum phenomena, including superconductivity \cite{Imada_1, Peotta}, ferromagnetism \cite{Mielke, Tasaki} and fractional quantum Hall effect \cite{Tang, DasSarma, Neupert, ShengDN}. Nonetheless, electronic-correlation-induced flat bands have rarely been observed in WSMs. Lately, single-particle \textit{ab initio} predictions of WSM states in 3\textit{d}-transition-metal compounds shed light on searching for correlated WSMs with flat bands, owing to the intimate association between the weak spatial extension of 3\textit{d} orbitals and large Coulomb interactions \cite{Qiunan Xu, Enke Liu, Hechang Lei, Yanghao}.   
   
A cobalt-based shandite compound, Co$_{3}$Sn$_{2}$S$_{2}$, crystallizes in a rhombohedral structure with the cobalt atoms forming a kagome lattice within one quasi-two-dimensional Co$_{3}$Sn layer (see Fig. 1a) and exhibits long-range ferromagnetic (FM) order with a magnetic moment of $\sim$ 0.3 $\rm {\mu_B}$ ($\rm \mu_B$ denotes the Bohr magneton) per cobalt atom below temperature \textit{T} $\sim$ 177 K \cite{Weihrich1, Weihrich2, Vaqueiro, Schnelle}. Single-particle \textit{ab initio} calculations show that the electronic bands of FM Co$_{3}$Sn$_{2}$S$_{2}$ near the Fermi energy ($E_{\rm F}$) are dominated by cobalt 3\textit{d} orbitals (see the electron density of states (DOS) for Co$_{3}$Sn$_{2}$S$_{2}$ with the Co 3\textit{d}, Sn 5\textit{p} and S 3\textit{p} orbital contributions shaded in red, green and purple colors, respectively in Fig. 1b) \cite{JXYin}, but the strength of electronic correlations in this FM 3\textit{d}-transition-metal compound remains unclear. Furthermore, single-particle \textit{ab initio} calculations suggest that FM Co$_{3}$Sn$_{2}$S$_{2}$ is a contender for magnetic WSMs \cite{Qiunan Xu, Enke Liu, Hechang Lei}. Up to now, important progresses in the experimental studies of the predicted WSM state in FM Co$_{3}$Sn$_{2}$S$_{2}$, which involve the measurements of negative magnetoresistance, giant intrinsic anomalous Hall, Nernst effects, bulk Weyl cones and surface Fermi arcs \cite{Enke Liu, Hechang Lei, Guin, YKLi, Jiao, Beidenkopf, Guguchia, YLChen_css, Howlader}, have been achieved. However, the influence of electronic correlations on the single-particle-\textit{ab-initio}-calculation-derived WSM state in this WSM candidate, for example, inducing a flat band, remains elusive. 
 
\vspace{3mm}
\noindent \textbf{Results} 
          
\noindent \textbf{Reduction of the electronic kinetic energy.} Optical spectroscopy is a bulk-sensitive experimental technique for studying charge dynamics and electronic band structure of a material as it probes both itinerant charge carriers and interband transitions from occupied to empty states \cite{Basov3, Basov2, Basov1, QMSi, NLWang, Martin Dressel}. Here, to investigate electronic correlations and their effects on the previously predicted WSM state in FM Co$_{3}$Sn$_{2}$S$_{2}$, we measured the optical reflectance spectra \textit{R}($\omega$) of its single crystals at low temperatures with the electric field (\textit{E}) of the incident light parallel to the crystalline \textit{ab}-plane over a broad photon energy ($\omega$) range (see the details about the reflectance measurements and the sample growth in Methods section). Figure 1c depicts the \textit{R}($\omega$) of Co$_{3}$Sn$_{2}$S$_{2}$ single crystals measured at different temperatures. The \textit{R}($\omega$) at energies lower than 20 meV not only approach to unity, but also increase as the temperature decreases, which exhibits the optical response of a metal. Moreover, the real parts (i.e., $\sigma_{1}$(\emph{$\omega$})) of the \textit{ab}-plane optical conductivity of Co$_{3}$Sn$_{2}$S$_{2}$ in Fig. 1d, which were obtained by the Kramers-Kronig transformation of the \textit{R}($\omega$) (see Methods section), show Drude-like features of metals at energies lower than 20 meV. The Drude-like features in the low-energy parts of the $\sigma_{1}$(\emph{$\omega$}) and the fast-increasing value of the \textit{R}($\omega$) at low energies indicate the existence of itinerant charge carriers in Co$_{3}$Sn$_{2}$S$_{2}$, which provides an opportunity for studying the electronic correlation effect on the motion of the itinerant charge carriers. Furthermore, several peak-like features arising from interband transitions are present in the high-energy parts of the $\sigma_{1}$(\emph{$\omega$}) (please see the possible relation between the decrease in the intensity of the peak-like feature and the absence of WSM phase in the paramagnetic (PM) state of this system in Supplementary Note 1). Comparing the energies and shapes of the experimental interband-transition-induced peak-like features with those of the peak-like features calculated without considering electronic correlations enables us to gain insights into the effect of electronic correlations on the bandwidth.  

\begin{figure*}
	\includegraphics[width=15.7cm]{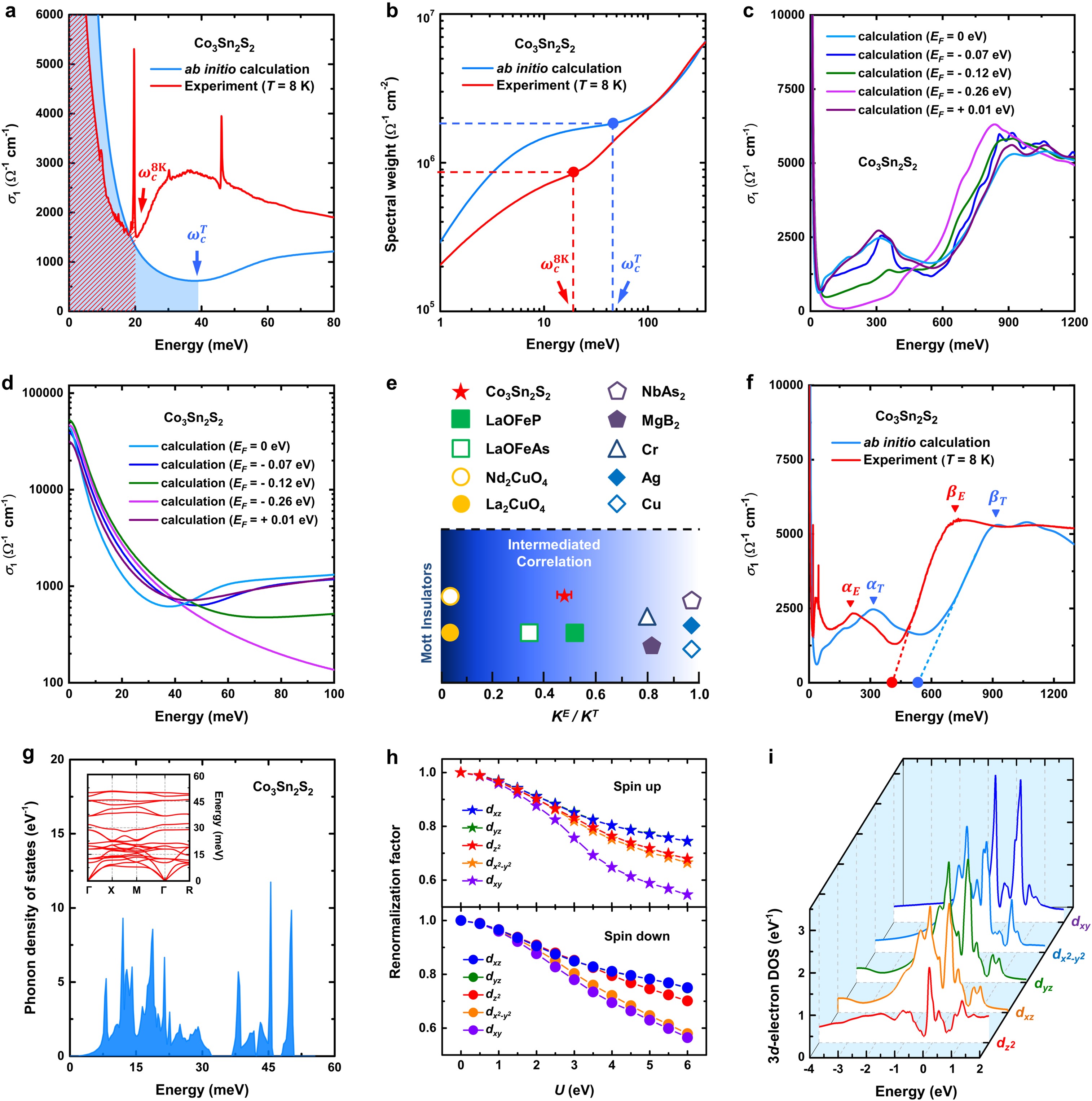}\\
	\centering
	\caption{\textbf{Electronic correlation effects in ferromagnetic Co$_{3}$Sn$_{2}$S$_{2}$.} \textbf{a}, Real parts of the experimental optical conductivity $\sigma_{1}^{\rm E}$(\emph{$\omega$}) at temperature \textit{T} = 8 K and the theoretical optical conductivity $\sigma_{1}^{\rm T}$(\emph{$\omega$}) obtained by single-particle \textit{ab initio} calculations in the FM ground state. The cut-off energies ($\omega_{\rm c}^{8\rm K}$ and $\omega_{\rm c}^{\rm T}$) for integrating the Drude components are chosen as the energy positions where the $\sigma_{1}^{\rm E}$(\emph{$\omega$}, \textit{T} = 8 K) and the $\sigma_{1}^{\rm T}$(\emph{$\omega$}) reach the minimum, respectively. \textbf{b}, Spectral weights of the $\sigma_{1}^{\rm E}$(\emph{$\omega$}, \textit{T} = 8 K) at $\omega_{\rm c}^{8\rm K}$ and the $\sigma_{1}^{\rm T}$(\emph{$\omega$}) at $\omega_{\rm c}^{\rm T}$. \textbf{c}, Theoretical $\sigma_{1}^{\rm T}$(\emph{$\omega$}) calculated with the Fermi energies $E_{\rm F}$ = 0 eV, -0.07 eV, -0.12 eV, -0.26 eV and + 0.01 eV. \textbf{d}, Low-energy parts of the theoretical $\sigma_{1}^{\rm T}$(\emph{$\omega$}) calculated with the different Fermi energies. \textbf{e}, Ratio of the experimental kinetic energy at \textit{T} = 8 K and theoretical kinetic energy $K_{8\rm K}^{\rm E}$/$K^{\rm T}$ for Co$_{3}$Sn$_{2}$S$_{2}$ and several other quantum materials. The values of the $K^{\rm E}$ and $K^{\rm T}$ for other quantum materials can be gained from the following references: LaOFeP (Ref. \cite{Basov1}), LaOFeAs (Ref. \cite{NLWang}), topological nodal-line semimetal NbAs$_2$ (Ref. \cite{Basov4}), paramagnetic Cr (Ref. \cite{van der Marel}), Ag (Ref. \cite{Romaniello2}), Cu (Ref. \cite{Romaniello2}), MgB$_2$ (Ref. \cite{Guritanu}), Nd$_2$CuO$_4$ and La$_2$CuO$_4$ (Ref. \cite{Millis}). The error bar on the Co$_{3}$Sn$_{2}$S$_{2}$ data is based on the uncertainty in the experimental cut-off energy $\omega_{\rm c}$. \textbf{f}, Two interband-transition-induced peaks $\alpha_{\rm E}$ and $\beta_{\rm E}$ in the $\sigma_{1}^{\rm E}$(\emph{$\omega$}, \textit{T} = 8 K) and the two peaks $\alpha_{\rm T}$ and $\beta_{\rm T}$ in the $\sigma_{1}^{\rm T}$(\emph{$\omega$}). The red and blue arrows indicate the left sides of the experimental peak $\beta_{\rm E}$ and the theoretical peak $\beta_{\rm T}$, respectively. The red and blue dashed lines are guides for eyes showing the slopes of the left sides of the peaks $\beta_{\rm E}$ and $\beta_{\rm T}$, respectively. The slope of the left side of the experimental peak $\beta_{\rm E}$ is larger than that of the theoretical peak $\beta_{\rm T}$. The ratio between the slopes of the left sides of the peaks $\beta_{\rm T}$ and $\beta_{\rm E}$ is $\sim$ 0.74. The energy intercepts of the red and dashed lines at $\sigma_{1}$(\emph{$\omega$}) = 0 are labelled with the red and blue dots, respectively. \textbf{g}, Phonon density of states obtained by \textit{ab initio} calculations. The inset depicts the calculated phonon dispersions. The phonon DOS and the phonon dispersions are cut off at $\sim$ 50.8 meV. \textbf{h}, Coulomb-energy dependences of the electronic-bandwidth renormalization factors of the spin-up and spin-down Co 3\textit{d} orbitals. The renormalization factors of the $d_{xz}$ and $d_{yz}$ orbitals are quite close to each other. \textbf{i}, Energy distributions of the 3\textit{d}-electron density of states obtained by many-body calculations. }
	\label{FIG.1.linear}  
\end{figure*}  

To study the electronic correlation effect on the motion of the itinerant charge carriers, we compare the experimentally measured kinetic energy with the theoretical kinetic energy calculated without taking any many-body interaction into account. Following the definition of the electronic kinetic energy in the optical study of a multiband system LaOFeP \cite{Basov1}, we can obtain the linear relationship between the electronic kinetic energy (\textit{K}) and the spectral weight ($S$) of the Drude component (i.e., the area under the Drude component) of the $\sigma_{1}$(\emph{$\omega$}):
\begin{eqnarray}
K = \frac{2\emph{$\hbar$}^{2}\emph{$d_0$}}{\pi\emph{e}^{2}}S = \frac{2\emph{$\hbar$}^{2}\emph{$d_0$}}{\pi\emph{e}^{2}}\int_{0}^{\omega_c}\sigma_{1}(\omega)\rm d \omega,
\end{eqnarray} 
where $\omega_{\rm c}$ is a cut-off energy for integrating the Drude component of the $\sigma_{1}$(\emph{$\omega$}), $\hbar$ is Planck's constant divided by 2$\pi$, \textit{e} is the elementary charge and $d_0$ is the inter-Co$_3$Sn-layer distance. Figure 2a displays the Drude components of the real part of the experimental optical conductivity $\sigma_{1}^{\rm E}$(\emph{$\omega$}) at \textit{T} = 8 K and the real part of the theoretical optical conductivity $\sigma_{1}^{\rm T}$(\emph{$\omega$}) obtained by single-particle \textit{ab initio} calculations of the FM ground state of Co$_{3}$Sn$_{2}$S$_{2}$ (see the Drude components over a broader range of the $\sigma_{1}$(\emph{$\omega$}) in Supplementary Fig. 2a). The cut-off energy $\omega_{\rm c}$ is usually chosen as the energy position where $\sigma_{1}$(\emph{$\omega$}) reaches its minimum below the interband transition, so the $\sigma_{1}^{\rm E}$(\emph{$\omega$}, \textit{T} = 8 K) and the $\sigma_{1}^{\rm T}$(\emph{$\omega$}) here have the cut-off energies $\omega_{\rm c}^{8\rm K}$ $\approx$ 19.9 $\pm$ 4 meV and $\omega_{\rm c}^{\rm T}$ $\approx$ 38.9 meV, respectively. Integrating the Drude components of the $\sigma_{1}^{\rm E}$(\emph{$\omega$}, \textit{T} = 8 K) and the $\sigma_{1}^{\rm T}$(\emph{$\omega$}) up to the cut-off energies $\omega_{\rm c}^{8\rm K}$ $\approx$ 19.9 $\pm$ 4 meV and $\omega_{\rm c}^{\rm T}$ $\approx$ 38.9 meV yields approximately the spectral weights of the experimental and theoretical Drude components: $S^{8\rm K}$ $\approx$ (8.6 $\pm$ 0.6) $\times$ $10^{5}$ $\Omega^{-1}$ cm$^{-2}$ and $S^{\rm T}$ $\approx$ 1.8 $\times$ $10^{6}$ $\Omega^{-1}$ cm$^{-2}$, respectively (see the red and blue points in Fig. 2b, the details about calculating the theoretical Drude spectral weight in the PM state $S^{\rm T}$ $\approx$ 4.0 $\times$ $10^{6}$ $\Omega^{-1}$ cm$^{-2}$ in Methods section and the theoretical Drude component in the PM state in Supplementary Fig. 3), which is consistent with the smaller area under the experimental Drude component compared with that under the calculated Drude component (see the red shaded area and the blue area in Supplementary Fig. 2a). Here, the theoretical Drude spectral weight in the FM ground state $S^{\rm T}$ $\approx$ 1.8 $\times$ $10^{6}$ $\Omega^{-1}$ cm$^{-2}$ is not impacted by the choice of the theoretical scattering rate which can significantly influence the cut-off energy (see Supplementary Note 2). Considering the linear relationship between the \textit{K} and the $S$, which is shown in Equation (1), we get the ratio between the experimental kinetic energy at \textit{T} = 8 K and the theoretical kinetic energy: $K_{8\rm K}^{\rm E}$/$K^{\rm T}$ = $S^{8\rm K}$/$S^{\rm T}$ $\approx$ 0.47 $\pm$ 0.04. To check the ratio $K_{8\rm K}^{\rm E}$/$K^{\rm T}$ between the experimental and theoretical kinetic energies, an alternative method based on the linear relationship between the kinetic energy and the square $\omega_{\rm D}^2$ of the plasma energy can be employed \cite{Basov3, Basov2, NLWang}. By fitting the experimental $\sigma_{1}^{\rm E}$(\emph{$\omega$}, \textit{T} = 8 K) of Co$_{3}$Sn$_{2}$S$_{2}$ based on a standard Drude-Lorentz model, we can obtain the experimental plasma frequency at \textit{T} = 8 K in its FM state: $\omega_{\rm D}^{\rm E}$ = 258 $\pm$ 4 meV (see the details in Methods section). Furthermore, the theoretical plasma energy $\omega_{\rm D}^{\rm T}$ of FM Co$_{3}$Sn$_{2}$S$_{2}$ can be directly calculated from the single-particle-\textit{ab-initio}-calculation-derived band structure, i.e., $\omega_{\rm D}^{\rm T}$ = 379 meV. Given the linear relationship between the kinetic energy and the square $\omega_{\rm D}^2$ of the plasma frequency $\omega_{\rm D}$, we can get the ratio between the experimental kinetic energy at \textit{T} = 8 K and the theoretical kinetic energy in the FM ground state: $K_{8\rm K}^{\rm E}$/$K^{\rm T}$ = ($\omega_{\rm D}^{\rm E}$/$\omega_{\rm D}^{\rm T}$)$^2$ $\approx$ 0.46 $\pm$ 0.02, which is consistent with the kinetic-energy ratio $K_{8\rm K}^{\rm E}$/$K^{\rm T}$ = $S^{8\rm K}$/$S^{\rm T}$ $\approx$ 0.47 $\pm$ 0.04 inferred from ratio between the integrations of the $\sigma_{1}^{\rm E}$(\emph{$\omega$}, \textit{T} = 8 K) and the $\sigma_{1}^{\rm T}$(\emph{$\omega$}) in the FM ground state up to the cut-off energies. Therefore, the ratios $K_{8\rm K}^{\rm E}$/$K^{\rm T}$ deduced by the above two methods indicate that the experimental kinetic energy of FM Co$_{3}$Sn$_{2}$S$_{2}$ at \textit{T} = 8 K is significantly smaller than the theoretical kinetic energy obtained by single-particle \textit{ab initio} calculations of FM Co$_{3}$Sn$_{2}$S$_{2}$. 

\begin{table}[!b]
	\begin{centering} 
		\caption{Ratios between the theoretical Drude weights $S^{\rm T}$($E_{\rm F}$) calculated with the different Fermi energies and the theoretical Drude weight $S^{\rm T}$($E_{\rm F}$ = 0 eV).}\label{tab:latticeparas} 
		\setlength{\tabcolsep}{2.05mm}{
			\begin{tabular}{|c|c|c|c|c|ccc}
				\hline 
				$E_{\rm F}$ (eV) & + 0.01 & -0.07 & -0.12 & -0.26 \\  
				\hline
				$S^{\rm T}$($E_{\rm F}$)/$S^{\rm T}$($E_{\rm F}$ = 0 eV) & 1.03 & 1.27 & 1.72 & 1.52 \\
				\hline
		\end{tabular}}
		\par\end{centering}
\end{table}

To check whether the substantial reduction of the experimental kinetic energy (or experimental Drude weight) compared with the theoretical kinetic energy (or theoretical Drude weight) can arise from the change in the Fermi level of FM Co$_{3}$Sn$_{2}$S$_{2}$, we performed single-particle \textit{ab initio} calculations of the $\sigma_{1}^{\rm T}$(\emph{$\omega$}) with different $E_{\rm F}$. The above theoretical $\sigma_{1}^{\rm T}$(\emph{$\omega$}) in Fig. 2a was obtained with $E_{\rm F}$ = 0 eV.  When $E_{\rm F}$ = 0 eV, the corresponding Fermi level is located at $\sim$ 0.06 eV below the calculated Weyl point \cite{Qiunan Xu, Enke Liu, Hechang Lei}, while the Fermi level measured by angle-resolved photoemission spectroscopy (ARPES) is located at $\sim$ 0.05 eV below the Weyl point \cite{YLChen_css}. Thus, the Fermi level measured by ARPES is $\sim$ 0.01 eV higher than the theoretical one corresponding to $E_{\rm F}$ = 0 eV. Figure 2c shows the theoretical $\sigma_{1}^{\rm T}$(\emph{$\omega$}) calculated with $E_{\rm F}$ = 0.01 eV in the energy range up to 1200 meV. The low-energy parts of the $\sigma_{1}^{\rm T}$(\emph{$\omega$}, $E_{\rm F}$ = 0.01 eV) and $\sigma_{1}^{\rm T}$(\emph{$\omega$}, $E_{\rm F}$ = 0 eV) in Fig. 2d indicate that the theoretical Drude weight $S^{\rm T}$($E_{\rm F}$ = 0.01 eV) is larger than the theoretical Drude weight $S^{\rm T}$($E_{\rm F}$ = 0 eV). As listed in Table 1, the ratio between the calculated $S^{\rm T}$($E_{\rm F}$ = 0.01 eV) and $S^{\rm T}$($E_{\rm F}$ = 0 eV) is $\sim$ 1.03. Thus, if the Fermi level is shifted up by 0.01 eV, the corresponding theoretical Drude weight $S^{\rm T}$($E_{\rm F}$ = 0.01 eV) will be larger than the $S^{\rm T}$($E_{\rm F}$ = 0 eV), which means that the theoretical kinetic energy at $E_{\rm F}$ = 0.01 eV will be larger than the theoretical one at $E_{\rm F}$ = 0 eV. In addition, we calculated the $\sigma_{1}^{\rm T}$(\emph{$\omega$}) with the negative Fermi energies $E_{\rm F}$ = -0.07 eV, -0.12 eV and -0.26 eV. Figure 2d, together with Table 1, shows that the theoretical Drude weights $S^{\rm T}$($E_{\rm F}$) corresponding to these negative Fermi energies are also larger than the theoretical Drude weight $S^{\rm T}$($E_{\rm F}$ = 0 eV), i.e., the theoretical kinetic energies at these negative Fermi energies are larger than the theoretical one at $E_{\rm F}$ = 0 eV as well. Therefore, the upshifting and lowering of the Fermi energy are unlikely to reduce the electronic kinetic energy of FM Co$_{3}$Sn$_{2}$S$_{2}$.

Figure 2e shows that the deduced ratios $K_{8\rm K}^{\rm E}$/$K^{\rm T}$ are distinctly smaller than unity---the kinetic-energy ratio in conventional metals (such as Ag and Cu) with quite weak effects of many-body interactions \cite{Romaniello2}. Here, the substantial reduction in the electronic kinetic energy compared with the $K^{\rm T}$ indicates that many-body interactions which have not been taken into account by single-particle \textit{ab initio} calculations in the FM ground state have a pronounced effect of impeding the motion of the itinerant charge carriers in FM Co$_{3}$Sn$_{2}$S$_{2}$. In contrast, ordered spin-spin correlations in itinerant ferromagnets usually correspond to an increase in the kinetic energy of itinerant charge carriers \cite{Blundell}, because (i) according to the Pauli exclusion principle, a larger kinetic energy is needed for the itinerant charge carriers with parallel spins to meet in the same lattice sites \cite{Herring}, and (ii) in the framework of the Stoner model, a phase transition from paramagnetism to itinerant ferromagnetism is accompanied with the increase in the electronic kinetic energy which is outweighed by the lowering of the exchange energy \cite{Stoner}. Thus, ordered spin-spin correlations in FM Co$_{3}$Sn$_{2}$S$_{2}$ are highly likely to be irrelevant with the remarkable reduction of the electronic kinetic energy here. Moreover, note that extremely strong electron-phonon coupling in a polar semiconductor or an ionic crystal can lead to a significant reduction of the electronic kinetic energy owing to the formation of polarons \cite{Kittel}. Nevertheless, the calculated cobalt-3\textit{d}-orbital-dominated bands which cross the $E_{\rm F}$ \cite{Qiunan Xu, Enke Liu, Hechang Lei, JXYin} and the measured magnetic moment ($\sim$ 0.3 $\rm \mu_B$/Co) which is much smaller than the magnetic moment (3 $\rm \mu_B$/Co) of isolated cobalt atoms \cite{Weihrich1, Weihrich2, Vaqueiro, Schnelle} strongly suggest that FM Co$_{3}$Sn$_{2}$S$_{2}$ should not be a polar semiconductor or an ionic crystal, either of which has been found to host polarons. Generally, electron-phonon coupling in a material with the absence of polarons would not make the ratio between experimental and theoretical kinetic energies much smaller than unity (see the $K^{\rm E}$/$K^{\rm T}$ in MgB$_2$ superconductor with electron-phonon mediated conventional superconductivity in Fig. 2e) \cite{Guritanu, Millis}, so for FM Co$_{3}$Sn$_{2}$S$_{2}$, electron-phonon coupling is also unlikely to be the main factor causing the substantial decrease in its electronic kinetic energy. Based on the above discussion, electronic correlations, which were previously revealed to result in the remarkable lowerings of the electronic kinetic energies in some transition-metal-based superconductors, such as the iron pnictides LaOFeP and LaOFeAs (see Fig. 2e) \cite{Basov1, NLWang}, should play a dominant role in hampering the motion of the itinerant charge carriers in FM Co$_{3}$Sn$_{2}$S$_{2}$. Since the $K_{8\rm K}^{\rm E}$/$K^{\rm T}$ in FM Co$_{3}$Sn$_{2}$S$_{2}$ is approximately equal to the average of the kinetic-energy ratio ($\sim$ 0) in Mott insulators (like Nd$_{2}$CuO$_{4}$ and Sr$_{2}$CuO$_{4}$) with very strong electronic correlations (see Fig. 2e) and the kinetic-energy ratio ($\sim$ 1) in conventional metals (such as Ag and Cu) with quite weak electronic correlations, the strength of electronic correlations in FM Co$_{3}$Sn$_{2}$S$_{2}$ can be regarded to be intermediate.  

\begin{figure*}
	\includegraphics[width=14.4cm]{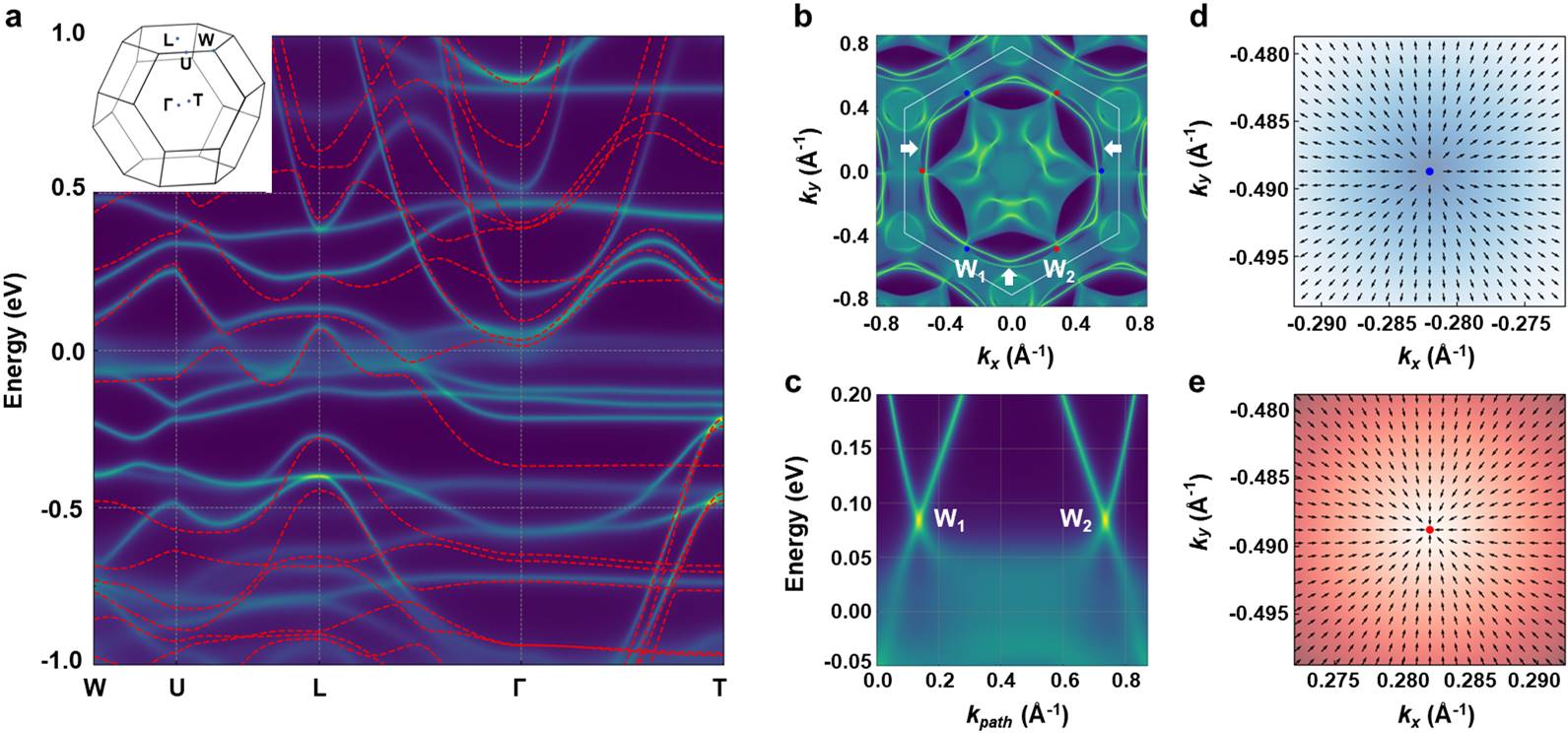}\\
	\centering  
	\caption{\textbf{Many-body-calculation-derived electronic structure of ferromagnetic Co$_{3}$Sn$_{2}$S$_{2}$.} \textbf{a}, Momentum-resolved electronic spectra and single-particle band-structure along the high-symmetry lines of the bulk Brillouin zone (the upleft inset). The momentum-resolved electronic spectra are in bright yellow color. The single-particle band-structure is represented by the red dashed curves. \textbf{b}, Fermi arcs on the (001) surface. The blue and red dots, W$_1$ and W$_2$, represent the projected Weyl points with positive and negative chirality, respectively. The three white arrows indicate the Fermi arcs. \textbf{c}, Bulk Weyl cones along the direction connecting the Weyl points W$_1$ and W$_2$ in \textbf{b}. \textbf{d}, \textbf{e}, Distribution of the Berry curvature around the Weyl points W$_1$ (\textbf{d}) and W$_1$ (\textbf{e}) in the $k_{xy}$ plane.}   
	\label{FIG.1.linear}    
\end{figure*}

\vspace{3mm}
\noindent \textbf{Narrowness of the electronic bandwidth.}
To investigate the effect of many-body interactions on the electronic bandwidth of FM Co$_{3}$Sn$_{2}$S$_{2}$, we plotted the $\sigma_{1}^{\rm E}$(\emph{$\omega$}, \textit{T} = 8 K) and the $\sigma_{1}^{\rm T}$(\emph{$\omega$}) over a broad energy range up to 1350 meV in Fig. 2f. The overall shape of the $\sigma_{1}^{\rm E}$(\emph{$\omega$}) at $\omega$ $>$ 20 meV is similar to that of the $\sigma_{1}^{\rm T}$(\emph{$\omega$}), but (i) the energy positions of the two peak-like features arising from interband transitions, $\alpha_{\rm E}$ at $\sim$ 217.4 meV and $\beta_{\rm E}$ at $\sim$ 708.1 meV, in the $\sigma_{1}^{\rm E}$(\emph{$\omega$}, \textit{T} = 8 K) are distinctly lower than those of the two corresponding peak-like features in the $\sigma_{1}^{\rm T}$(\emph{$\omega$}), $\alpha_{\rm T}$ at $\sim$ 319.7 meV and $\beta_{\rm T}$ at $\sim$ 931.6 meV, respectively; and (ii) the left side of the experimental peak-like feature $\beta_{\rm E}$ is significantly steeper than that of the theoretical peak $\beta_{\rm T}$ (the red and blue dashed lines in Fig. 2f are guides for eyes showing the slopes of the left sides of the peaks $\beta_{\rm E}$ and $\beta_{\rm T}$, respectively). Generally, when only the widths of the conduction and valence bands related to interband transitions are reduced, the width of the peak-like feature in $\sigma_{1}$(\emph{$\omega$}) arising from the interband transitions between these related bands decreases due to the reduced widths of these bands (see Supplementary Fig. 5). The reduction in the width of the peak-like feature in $\sigma_{1}$(\emph{$\omega$}), together with the unchanged height of the peak-like feature in $\sigma_{1}$(\emph{$\omega$}), would further lead to the increase in the slope of the sides of the peak-like feature (see the red dashed lines in see Supplementary Fig. 5b and 5d). Therefore, in Fig. 2f, the steeper left-side of the experimental peak $\beta_{\rm E}$ in the $\sigma_{1}^{{\rm E}}$(\emph{$\omega$}, \textit{T} = 8 K) of FM Co$_{3}$Sn$_{2}$S$_{2}$, combined with the comparability between the heights of the experimental and theoretical peaks $\beta_{\rm E}$ and $\beta_{\rm T}$, indicates the many-body-interaction-induced narrowing of the widths of the electronic bands in FM Co$_{3}$Sn$_{2}$S$_{2}$. In addition, the energy intercept of the red dashed line at the experimental $\sigma_{1}^{\rm E}$(\emph{$\omega$}, \textit{T} = 8 K) = 0, which can approximately represent the energy gap between the conduction and valence bands (see Supplementary Fig. 6) (or the minimal energy difference between the occupied and empty states in the electronic bands displayed in Supplementary Fig. 7), locates at lower energy than the energy intercept of the blue dashed line at the theoretical $\sigma_{1}^{\rm T}$(\emph{$\omega$}) = 0. This indicates that in FM Co$_{3}$Sn$_{2}$S$_{2}$, the experimental energy gap between the conduction and valence bands (or the experimental minimal energy difference between the occupied and empty states in the electronic bands) is smaller than the theoretical band gap (or the theoretical minimal energy difference). Besides, it is worth noticing that the change in the energy range (i.e., $\Delta$$\omega$) from the energy intercept of the dashed lines at $\sigma_{1}$(\emph{$\omega$}) = 0 to the energy position of the optical-conductivity-peak-height maximum can also reflect the renormalization of the widths of the electronic bands. Correspondingly, the ratio ($\sim$ 0.76) between the energy range $\Delta$$\omega$$_{\rm E}$ in the experimental $\sigma_{1}^{\rm E}$(\emph{$\omega$}, \textit{T} = 8 K) and the energy range $\Delta$$\omega$$_{\rm T}$ in the theoretical $\sigma_{1}^{\rm T}$(\emph{$\omega$}), which is comparable to the ratio ($\sim$ 0.74) between the slopes of the left sides of the peaks $\beta_{\rm E}$ and $\beta_{\rm T}$, indicates the narrowing of the widths of the electronic bands in FM Co$_{3}$Sn$_{2}$S$_{2}$ as well. Therefore, the red-shift of the interband-transition-induced peak $\beta_{\rm E}$ in the experimental $\sigma_{1}^{\rm E}$(\emph{$\omega$}, \textit{T} = 8 K) compared with the theoretical peak $\beta_{\rm T}$ in the calculated $\sigma_{1}^{\rm T}$(\emph{$\omega$}) not only indicates the decrease in the energy gap between the occupied and empty band (or the minimal energy difference between the occupied and empty states in the electronic bands), but also reflects the many-body-interaction-induced narrowing of the widths of the electronic bands in FM Co$_{3}$Sn$_{2}$S$_{2}$.

Given that (i) according to the Pauli exclusion principle and the Stoner model, ferromagnetically ordered spin-spin correlations usually leads to a gain in the electronic kinetic energy \cite{Blundell, Herring, Stoner}, (ii) the gain in the electronic kinetic energy mostly corresponds to an extension of the electronic bandwidth \cite{Blundell, Herring, Stoner}, and (iii) the cut-off energy ($\sim$ 50.8 meV) of the phonon spectrum shown in Fig. 2g is much lower than the energies of the interband-transition-induced peaks, $\alpha_{\rm E}$, $\alpha_{\rm T}$, $\beta_{\rm E}$, and $\beta_{\rm T}$, ferromagnetically ordered spin-spin correlations and electron-phonon coupling are unlikely to be the leading interactions which cause the narrowing of the electronic bandwidth here. Therefore, electronic correlations in FM Co$_{3}$Sn$_{2}$S$_{2}$ ought to play a major part in narrowing the electronic bandwidth. 

To estimate the Coulomb-interaction strength \textit{U} of electronic correlations in FM Co$_{3}$Sn$_{2}$S$_{2}$, we performed many-body calculations, i.e., combination of density functional theory and
dynamical mean-field theory (DFT+DMFT) (see the details in Methods section) \cite{Kotliar2, Kotliar3}, and then
obtained the \textit{U} dependences of the electronic-bandwidth renormalization factor quantifying the effect of electronic correlations on narrowing the electronic bandwidth (see Fig. 2h). The ratio between the slopes of the left sides of the experimental and theoretical peaks, \textit{S}($\beta_{\rm T}$)/\textit{S}($\beta_{\rm E}$) $\approx$ 0.74 and the energy ratios between the experimental and theoretical peaks in Fig. 2f, \textit{E}($\alpha_{\rm E}$)/\textit{E}($\alpha_{\rm T}$) $\approx$ 0.68 and  \textit{E}($\beta_{\rm E}$)/\textit{E}($\beta_{\rm T}$) $\approx$ 0.76, which are comparable to the ratio ($\sim$ 0.70) between the recently measured bandwidth and the calculated bandwidth \cite{YLChen_css}, reflect the electronic-bandwidth renormalization effect of the electronic correlations with \textit{U} $\sim$ 4 eV, shown in Fig. 2h. Here, the difference in the two energy ratios may arise from the discrepancy between the renormalization factors of the five 3\textit{d} orbitals (see Fig. 2h) and the difference between the energy distributions of the 3\textit{d}-electron DOS gotten by DFT+DMFT calculations (see Fig. 2i).

\begin{figure*}
	\includegraphics[width=15.8cm]{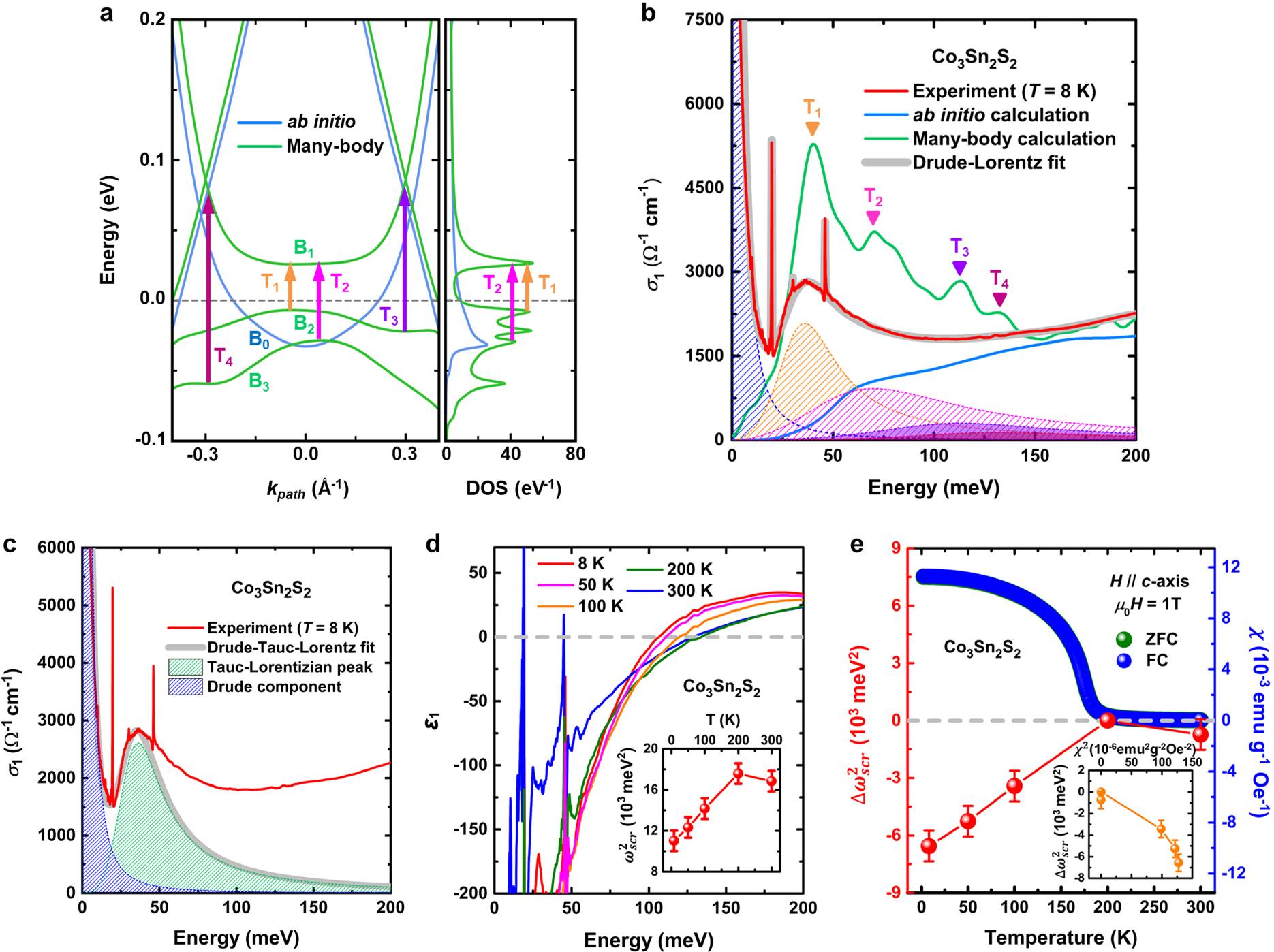}\\
	\centering       
	\caption{\textbf{Flat band and related optical transitions in ferromagnetic Co$_{3}$Sn$_{2}$S$_{2}$.} \textbf{a}, Left panel: quasiparticle band structure obtained by many-body calculations (i.e., DFT+DMFT calculations) (green curve) and band structure gotten by single-particle \textit{ab initio} calculations in the FM ground state (blue curve) along the direction connecting the Weyl points. Right panel: density of states (DOS) of the quasiparticle bands. The four arrows in the left panel show the optical transitions related to the flat band B$_1$ or the dispersionless part of band B$_2$ (or B$_3$) with divergent DOS. \textbf{b}, Experimental and calculated real parts $\sigma_{1}$(\emph{$\omega$}) of the optical conductivity at low energies. Four peak-like features (T$_1$, T$_2$, T$_3$ and T$_4$) in the $\sigma_{1}$(\emph{$\omega$}) obtained by many-body calculations are present around 39 meV, 70 meV, 113 meV and 131 meV, respectively. These four peak-like features arise mainly from the four optical  transitions illustrated by the four arrows in \textbf{a}. The asymmetric peak-like feature around 36 meV in the experimental $\sigma_{1}$(\emph{$\omega$}) at \textit{T} = 8 K can be fitted with the four Lorentzian peaks (see the shaded peaks in \textbf{b}), which are located around 36 meV, 70 meV, 113 meV and 131 meV, respectively. Peak-like feature is absent around 38 meV in the interband-transition-contributed part of the  $\sigma_{1}^{\rm T}$(\emph{$\omega$}) obtained by single-particle \textit{ab initio} calculations in the FM ground state (see the blue spectrum in \textbf{b}). \textbf{c}, Drude-Tauc-Lorentz fit to the experimental $\sigma_{1}^{\rm E}$(\emph{$\omega$}, \textit{T} = 8 K) of Co$_{3}$Sn$_{2}$S$_{2}$. The green and blue shaded areas are the Tauc-Lorentzian peak and the Drude component, respectively. The gray curve shows the Drude-Tauc-Lorentz fit to the experimental peak-like feature around 36 meV in the $\sigma_{1}^{\rm E}$(\emph{$\omega$}, \textit{T} = 8 K). \textbf{d}, Real parts $\varepsilon_1$(\emph{$\omega$}) of the dielectric functions of Co$_{3}$Sn$_{2}$S$_{2}$ at different temperatures. The inset of \textbf{d} shows the squares $\omega_{\rm scr}^2$ of its screened plasma frequencies at different temperatures. \textbf{e}, Relative squares of the screened plasma frequency $\Delta$$\omega_{\rm scr}^2$ = $\omega_{\rm scr}^2$(\textit{T}) - $\omega_{\rm scr}^2$(\textit{T} = 200 K) at different temperatures (see red dots) and temperature dependence of the magnetic susceptibility $\chi$ with zero-field-cool (ZFC) and field-cool (FC) modes in a magnetic field \textit{B} = 1 T for \textit{B}//\textit{c}-axis (see green and blue dots). In \textbf{e}, the ZFC-mode susceptibility data are covered by the FC-mode susceptibility data \cite{Hechang Lei}. The inset of \textbf{e} displays the $\chi^2$ dependence of the $\Delta$$\omega_{\rm scr}^2$. The error bars on the data in \textbf{d} and \textbf{e} are based on the line thickness of the $\varepsilon_1$(\emph{$\omega$}) in \textbf{d}.}           
	\label{FIG.1.linear}    
\end{figure*}  

\vspace{3mm}
\noindent \textbf{Persistence of a Weyl semimetal state.} To check whether a WSM state still exist in correlated Co$_{3}$Sn$_{2}$S$_{2}$, we carried out DFT+DMFT calculations with \textit{U} $\sim$ 4 eV to obtain its electronic surface and bulk states (see Methods section). In Fig. 3a, compared with the \textit{ab-initio}-calculation-derived bulk bands (see the red dashed curves) along the high-symmetry lines of the Brillouin zone (see the upleft inset), the bulk momentum-resolved electronic spectra (see the bright yellow curves) gotten by DFT+DMFT calculations are indeed renormalized. In Fig. 3b, the Fermi-arc structures on the (001) surface, which are based on the quasiparticle bands from DFT+DMFT calculations, connect three pairs of Weyl points, respectively. Figure 3c depicts a pair of bulk Weyl cones along the direction (i.e., W$_1$-W$_2$) connecting the Weyl points W$_1$ (i.e., blue point) and W$_2$ (i.e., red point) in Fig. 3b. To study the chirality of these two Weyl points W$_1$ and W$_2$, we calculated the Berry curvature around each Weyl point. As displayed in Fig. 3d and 3e, W$_1$ and W$_2$ act as a source and a sink of Berry curvature, respectively, so W$_1$ and W$_2$ have opposite chirality \cite{Xiangang Wan, A. A. Burkov, Gang Xu, Hongming Weng, Shin-Ming Huang, Soluyanov}. The surface Fermi arc connecting each pair of Weyl points with opposite chirality and the bulk Weyl cones in FM Co$_{3}$Sn$_{2}$S$_{2}$, which were obtained by our DFT+DMFT calculations, indicate the existence of a WSM state in this magnetic system with intermediate-strength electronic correlations. 
  
\vspace{3mm} 
\noindent \textbf{Flat band connecting the two Weyl cones.} To further search for possible effects of intermediate-strength electronic correlations on the WSM state in FM Co$_{3}$Sn$_{2}$S$_{2}$, we derived its quasiparticle band structure along the direction W$_1$-W$_2$ connecting the two Weyl points via DFT+DMFT calculations. The left panel of Fig. 4a shows that (i) a band B$_0$ obtained by single-particle \textit{ab initio} calculations, which not only is a part of the two Weyl cones but also links the two Weyl cones, is turned into a flat band B$_1$ near $E_{\rm F}$ in the quasiparticle band structure by electronic correlations, and that (ii) two bands B$_2$ and B$_3$, which have dispersionless parts and are absent along W$_1$-W$_2$ in the single-particle band structure, emerges below $E_{\rm F}$ in the quasiparticle band structure. Since (i) the flat band B$_1$ and the dispersionless parts of bands B$_2$ and B$_3$ have divergent DOS (see the right panel of Fig. 4a) and (ii) optical absorptions are determined by the joint DOS of the initial and final state, the four interband transitions related to the flat band B$_1$ or the dispersionless part of band B$_2$ (or B$_3$), which are illustrated by the four colored arrows in the left panel of Fig. 4a, cause the four obvious peak-like features T$_1$, T$_2$, T$_3$ and T$_4$ around 38 meV, 70 meV, 113 meV and 131 meV in the real part of the optical conductivity  $\sigma_{1}^{\rm QP}$(\emph{$\omega$}) contributed by the direct optical transitions between the calculated quasiparticle bands, respectively (see the green spectrum Fig. 4b and the details on calculating the $\sigma_{1}^{\rm QP}$(\emph{$\omega$}) in Methods section). Therein, (i) the strongest peak-like feature T$_1$ comes primarily from the optical transitions between the flat bands B$_1$ and the top of band B$_2$ (see the yellow arrow in Fig. 4a), and (ii) the second strongest peak-like feature T$_2$ arises mainly from the optical transitions between the flat band B$_1$ and the top of band B$_3$ (see the pink arrow in Fig. 4a). Considering that the present of the peak-like features T$_1$ and T$_2$ are intimately associated with the existence of the flat band B$_1$, the peak-like features T$_1$ around 38 meV and T$_2$ around 70 meV can be regarded as two spectroscopic signatures of the existence of the flat band B$_1$. It is worth noticing that the experimental $\sigma_{1}^{\rm E}$(\emph{$\omega$}, \textit{T} = 8 K) of FM Co$_{3}$Sn$_{2}$S$_{2}$ has an asymmetric peak-like feature around 36 meV, which cannot be well reproduced by only one Lorentzian term in a standard Drude-Lorentz model or a single Tauc-Lorentzian term in the Drude-Tauc-Lorentz model (see the Drude-Tauc-Lorentz fit to the asymmetric peak-like feature around 36 meV in Fig. 4c and the details about the Drude-Tauc-Lorentz fit in Methods section) \cite{Basov3, Basov2, Basov1, QMSi, NLWang, Martin Dressel, Jellison}. By fitting the low-energy part of the  $\sigma_{1}^{\rm E}$(\emph{$\omega$}, \textit{T} = 8 K) based on the Drude-Lorentz model, we find that this experimental peak-like feature can be decomposed into four components: a Lorentzian peak with the strongest intensity around 36 meV, a Lorentzian peak with the second strongest intensity around 70 meV, a Lorentzian peak around 113 meV and a Lorentzian peak around 131 meV  (see Methods section, the gray spectrum and the shaded peaks in Fig. 4b), which are consistent with the four peak-like features originating from the optical transitions related to the flat band B$_1$ or the dispersionless part of band B$_2$ (or B$_3$). Besides, this asymmetric peak-like feature in the experimental  $\sigma_{1}^{\rm E}$(\emph{$\omega$}, \textit{T} = 8 K) becomes weaker as the temperature increases and disappears completely above the FM transition temperature (i.e., not in the WSM state) (see Fig. 1d), which is in agreement with the absence of the Weyl cones and the flat band connecting the Weyl cones above the FM transition temperature. Therefore, the experimental peak-like feature around 36 meV in the  $\sigma_{1}^{\rm E}$(\emph{$\omega$}, \textit{T} = 8 K), which is obvious only in the FM state at low temperatures (i.e., in the WSM state) and includes the two Lorentzian peaks around the energy positions of T$_1$ and T$_2$, provides spectroscopic evidence for the existence of the flat band B$_1$ near $E_{\rm F}$ in FM Co$_{3}$Sn$_{2}$S$_{2}$. 

To check whether the experimental peak-like feature around 36 meV in the $\sigma_{1}^{\rm E}$(\emph{$\omega$}, \textit{T} = 8 K) could be explained by the exchange splitting of the double exchange model, we investigated the relationship between the relative square $\Delta$$\omega_{\rm scr}^2$ of the screened plasma frequency normalized to the $\omega_{\rm scr}^2$ at \textit{T} $\approx$ 200 K and the square $\chi^2$ of the magnetic susceptibility because a linear relationship between $\Delta$$\omega_{\rm scr}^2$ and $\chi^2$ was not only expected in the double exchange model but also observed in the manganites and FM semiconductors Ga$_{1-x}$Mn$_{x}$As \cite{Basov4, Furukawa, Okimoto, Noh, Ishikawa}. Figure 4d shows the real part $\varepsilon_{1}$(\emph{$\omega$}) of the dielectric function of Co$_{3}$Sn$_{2}$S$_{2}$ at different temperatures, which can be obtained from the imaginary part $\sigma_{2}$(\emph{$\omega$}) of its experimental \textit{ab}-plane optical conductivity. Since the screened plasma frequency $\omega_{\rm scr}$ is equal to the energy at which the real part of the dielectric function $\varepsilon_{1}$(\emph{$\omega$}) = 0, we can plot the $\omega_{\rm scr}^2$ as a function of temperature in the inset of Fig. 4d. The experimental $\omega_{\rm scr}^2$(\textit{T} = 8 K) in the FM state is smaller than the experimental $\omega_{\rm scr}^2$(\textit{T} = 200 K) in the PM state, which is consistent with the single-particle-\textit{ab-initio}-calculation-derived result that the theoretical $\omega_{\rm scr}^2$ in the FM ground state is lower than the theoretical one in the PM state (see Supplementary Fig. 3 and the details about the single-particle \textit{ab initio} calculations of the PM state in Methods section). Figure 4e displays the $\Delta$$\omega_{\rm scr}^2$ and the $\chi$ of Co$_{3}$Sn$_{2}$S$_{2}$ at different temperatures. As shown in the inset of Fig. 4e, the $\Delta$$\omega_{\rm scr}^2$ in the FM state of Co$_{3}$Sn$_{2}$S$_{2}$ does not exhibit a linear dependence on $\chi^2$, which is inconsistent with the linear relationship between the $\Delta$$\omega_{\rm scr}^2$ and the $\chi^2$ within the double exchange model. Therefore, it seems unlikely that the experimental peak-like feature around 36 meV in the $\sigma_{1}^{\rm E}$(\emph{$\omega$}, \textit{T} = 8 K) of FM Co$_{3}$Sn$_{2}$S$_{2}$ could arise from exchange-splitting-induced-interband transitions. 

In addition, single-particle \textit{ab initio} calculations indicate that the Weyl points of FM Co$_{3}$Sn$_{2}$S$_{2}$ are located at $\sim$ 60 meV above the Fermi level \cite{Qiunan Xu, Enke Liu, Hechang Lei}. According to Pauli's exclusion principle, the onset energy $E_{\rm onset}$ of the interband transitions between the occupied and empty states of the Weyl cones with the linear dispersions is about double the energy difference between the Weyl point and the Fermi level, i.e., $E_{\rm onset}$ $\sim$ 120 meV, which is much higher than the energy position ($\sim$ 36 meV) of the asymmetric peak-like feature in the $\sigma_{1}^{\rm E}$(\emph{$\omega$}, \textit{T} = 8 K). Thus, the asymmetric peak-like feature around 36 meV cannot originate from the interband transitions between the occupied and empty states of the Weyl cones with linear dispersions.

Moreover, in stark contrast to the experimental $\sigma_{1}^{\rm E}$(\emph{$\omega$}, \textit{T} = 8 K) and the calculated $\sigma_{1}^{\rm QP}$(\emph{$\omega$}), the interband-transition-contributed part of the  $\sigma_{1}^{\rm T}$(\emph{$\omega$}) obtained by single-particle \textit{ab initio} calculations in the FM ground state with the different Fermi energies has no distinct peak-like feature around 38 meV (see Fig. 2c and Fig. 4b), which further supports that electronic correlations in FM Co$_{3}$Sn$_{2}$S$_{2}$ flatten the band linking the two Weyl cones and induce the emergence of the flat band B$_1$.
 
\vspace{3mm}
\noindent \textbf{Discussion}
    
\noindent In summary, we have investigated electronic correlations in FM Co$_{3}$Sn$_{2}$S$_{2}$. The electronic kinetic energy extracted from the measured optical data is about half of that deduced by single-particle \textit{ab initio} calculations, which indicates that the strength of electronic correlations in Co$_{3}$Sn$_{2}$S$_{2}$ is intermediate. The energies of the two interband-transition-induced peaks in the experimental $\sigma_{1}^{\rm E}$(\emph{$\omega$}, \textit{T} = 8 K) of Co$_{3}$Sn$_{2}$S$_{2}$ are significantly lower than those in the $\sigma_{1}^{\rm T}$(\emph{$\omega$}) obtained by single-particle \textit{ab initio} calculations in the FM ground state. In addition, the left side of the experimental peak $\beta_{\rm E}$ is distinctly steeper than that of the theoretical peak $\beta_{\rm T}$. The red-shift and the steeper-side of the interband-transition-induced peak in the experimental $\sigma_{1}^{\rm E}$(\emph{$\omega$}, \textit{T} = 8 K) compared with the theoretical peak in the theoretical $\sigma_{1}^{\rm T}$(\emph{$\omega$}) indicate that its electronic bandwidth and band gap (or the minimal energy difference between the occupied and empty state in the electronic bands) are narrowed by electronic correlations. Furthermore, by comparing the energy ratios between the interband-transition peaks in the experimental and single-particle-\textit{ab-initio}-calculation-derived real parts of the optical conductivity with the electronic-bandwidth renormalization factors gotten by DFT+DMFT calculations, we estimated the Coulomb-interaction strength (\textit{U} $\sim$ 4 eV) of electronic correlations in this material. Our DFT+DMFT calculations with \textit{U} $\sim$ 4 eV show that a WSM state still exists in this correlated system. Besides, the consistence between the asymmetric peak-like feature around 36 meV in the experimental $\sigma_{1}^{\rm E}$(\emph{$\omega$}, \textit{T} = 8 K) and the DFT+DMFT-calculation-derived peak-like features in the $\sigma_{1}^{\rm QP}$(\emph{$\omega$}) reveals an electronic band connecting the two Weyl cones is flattened by electronic correlations and is present near $E_{\rm F}$ in FM Co$_{3}$Sn$_{2}$S$_{2}$. Our results not only reveal the effects of electronic correlations in FM Co$_{3}$Sn$_{2}$S$_{2}$, but also open an avenue for deeply investigating exotic quantum phenomena dominated by flat bands in WSMs. 
 
\vspace{3mm}
\noindent \textbf{Methods}

\vspace{1mm} 
\noindent \textbf{Optical reflectance measurements.} The optical reflectance measurements in the energy range from 8 to 6000 meV were performed on a Bruker Vertex 80v Fourier-transform spectrometer. The single-crystal sample was mounted on an optically black cone locating at the cold finger of a helium flow cryostat. A freshly-cleaved \textit{ab}-plane of the Co$_{3}$Sn$_{2}$S$_{2}$ single crystal was obtained just before pumping the cryostat. An \textit{in situ} gold and aluminum overcoating technique was employed to get the reflectance spectra \textit{R}($\omega$). The optical reflectance data are highly reproducible. Moreover, J. A. Woollam RC2 spectroscopic ellipsometer was used to get the optical constants of the Co$_{3}$Sn$_{2}$S$_{2}$ single crystals in the energy range from 500 to 6000 meV, which are consistent with the optical constants extracted from the measured reflectance spectra in this energy range.

\vspace{1mm} 
\noindent \textbf{Single-crystal growth.} The Co$_{3}$Sn$_{2}$S$_{2}$ single crystals were grown by a self-flux method. High-purity elemental Co, Sn and S with a molar ratio of 3:2:2 were put into an alumina crucible and then sealed in a quartz tube under high vacuum. The quartz tube was slowly heated to 637 K and maintained for two days due to the high vapor pressure of sulfur. Afterwards, the quartz tube was heated to 1273 K within 10 hours and then slowly cooled down to 973 K before switching off the furnace. Shining crystal faces can be obtained by cleaving the Co$_{3}$Sn$_{2}$S$_{2}$ single crystals.

\vspace{1mm} 
\noindent \textbf{Kramers-Kronig transformation.} The $\sigma_{1}$(\emph{$\omega$}) were obtained by the Kramers-Kronig transformation of the \textit{R}($\omega$). A Hagen-Rubens relation was used for low-energy extrapolation, and a $\omega^{-0.15}$ dependence was used for high-energy extrapolation up to 80000 meV, above which a $\omega^{-4}$ dependence was employed. The reciprocal value of the obtained $\sigma_{1}$(\emph{$\omega$} = 0) at each temperature coincides with the direct current resistivity obtained by the transport measurements (see Supplementary Fig. 2b), which indicates that the Kramers-Kronig transformation of the \textit{R}($\omega$) here is reliable.

\vspace{1mm} 
\noindent \textbf{Single-particle \textit{ab initio} calculations.} Our single-particle \textit{ab initio} optical conductivity calculations were performed at \textit{T} = 0 K in the FM and PM ground states of Co$_{3}$Sn$_{2}$S$_{2}$ by using the full potential linearized augmented plane wave method implemented in the WIEN2k package (see the spin-polarized bands in Supplementary Fig. 1b) \cite{wien2k}. The $k$-point mesh for the Brillouin zone integration is $36\times 36\times 36$, and the plane wave cut-off $K_{\rm max}$ is given by $R_{\rm mt}*K_{\rm max}=8.0$. The spin-orbit coupling effects are included in our calculations. The phonon dispersions were calculated by using the open source code PHONOPY \cite{phonopy}. The phonon force constants in real space were calculated based on the density-functional perturbation theory (DFPT) method using Vienna \textit{ab initio} simulation package (VASP) \cite{PhysRevB.54.11169} with a $2\times 2\times 2$ supercell. The plane wave energy cut-off was chosen as 400 eV, and a $\Gamma$-centered $k$-point grid with $3\times 3\times 3$ discretization was used.

Furthermore, we employed two different methods—HSE06 hybrid functional method in VASP package and mBJ method in WIEN2k package to calculate the electronic band structure of FM Co$_3$Sn$_2$S$_2$. The electronic band structures calculated by HSE06 hybrid functional method and mBJ method do not exhibit band inversions near the Fermi energy (please see the detailed results in Supplementary Note 3).

\vspace{1mm} 
\noindent \textbf{Fitting based on the Drude-Lorentz model.} We fit the $\sigma_{1}^{\rm E}$(\emph{$\omega$}, \textit{T} = 8 K) of Co$_3$Sn$_2$S$_2$ using a standard Drude-Lorentz model \cite{Basov3, Basov2, Basov1, QMSi, NLWang, Martin Dressel}:
\begin{eqnarray} 
\emph{$\sigma$}_{1}(\emph{$\omega$})= \frac{2\pi}{Z_0}\frac{\emph{$\omega$}^{2}_{\rm D}{\Gamma}_{\rm D}}{{\omega}^{2}+{\Gamma}^{2}_{\rm D}}+\sum_{j=1}^N  \frac{2\pi}{Z_0}\frac{{S}^{2}_{j}{\omega}^{2}{\Gamma}_{j}}{({\omega}_{j}^{2}-{\omega}^{2})^{2}+{\omega}^{2}{\Gamma}_{j}^{2}},
\end{eqnarray} 
where $Z_0$ $\approx$ 377 $\Omega$ is the impedance of free space, \emph{$\omega$}$\!_{\rm D}$ is the plasma frequency, and $\Gamma_{\rm D}$ is the relaxation rate of itinerant charge carriers, while $\omega_j$, $\Gamma_j$ and $S_j$ are the resonance frequency, the damping, and the mode strength of each Lorentz term, respectively.
The first term in Equation (2) denotes the optical response of free carriers, i.e., Drude response. The Lorentzian  terms can describe the contributions from inter--band transitions. The parameters of the four Lorentzian terms and the Drude term for fitting the low-energy part of the $\sigma_{1}^{\rm E}$(\emph{$\omega$}, \textit{T} = 8 K) are listed in Table 2.
\begin{table}[htbp] 
	\begin{centering} 
		\caption{Parameters of the Lorentzian and Drude terms.}\label{tab:latticeparas} 
		\begin{tabular}{cccccccc}
			\hline 
			& \textit{j} & \emph{$\omega$}$_{j}$ (meV) & \emph{$\Gamma$}$_{j}$ (meV) & \emph{S}$_{j}$ (meV) & \emph{$\omega$}$\!_{\rm D}$ (meV) & $\Gamma_{\rm D}$ (meV) \\  
			\hline
			& 1 & 36 & 37 & 214 & -- & -- \\
			& 2 & 70 & 98 & 231 & -- & -- \\   
			& 3 & 113 & 108 & 139 & -- & -- \\
			& 4 & 131 & 108 & 92 & -- & -- \\ 
			& D & -- & -- & -- & 258 & 2.5 \\
			\hline
		\end{tabular}
		\par\end{centering}
\end{table}

We further fit the low-energy part of the experimental $\sigma_{1}^{\rm E}$(\emph{$\omega$}, \textit{T} = 8 K) of Co$_3$Sn$_2$S$_2$ using a Drude-Tauc-Lorentz model. The Tauc-Lorentzian term for fitting can be expressed as \cite{Jellison}:
\begin{eqnarray} 
\emph{$\sigma$}_{1}^{\rm Tauc}(\emph{$\omega$}) \propto \frac{(\omega-E_{\rm g})^2}{({\omega}^2-E_{0}^2)^2+C^2{\omega}^2},
\end{eqnarray}
Here, $E_{\rm g}$ is the band gap, $E_0$ is the peak-transition energy and \textit{C} is the peak broadening term. The parameters of the Tauc-Lorentzian term for fitting are listed in Table 3.

\newcommand{\tabincell}[2]{\begin{tabular}{@{}#1@{}}#2\end{tabular}} 
\begin{table}[htbp] 
	\begin{centering} 
		\caption{Parameters of the Tauc-Lorentzian term.}\label{tab:latticeparas} 
		\begin{tabular}{cccccccc}
			\hline 
			& \tabincell{c}{Band gap\\$E_{\rm g}$ (meV)}  & \tabincell{c}{Peak-transition energy\\$E_0$ (meV)} & \tabincell{c}{Peak broadening\\ \textit{C} (meV)}  \\  
			\hline
			& 4 & 35 & 36 \\
			\hline
		\end{tabular}
		\par\end{centering}
\end{table}

\vspace{1mm} 
\noindent \textbf{Many-body calculations.} The method of density functional theory plus dynamical mean field theory (DFT+DMFT) can capture dynamic quantum fluctuation effects and thus is suitable for investigating the quasiparticles in correlated metals, while DFT+\textit{U} method is a static Hatree-Fock approach (see the band structures of Co$_3$Sn$_2$S$_2$ obtained by DFT+\textit{U} calculations in Supplementary Fig. 9 and Supplementary Note 4). The correlated electronic structure of Co$_3$Sn$_2$S$_2$ was obtained by DFT+DMFT calculations. A Wannier tight binding (TB) Hamiltonian consisting of $3d$ orbitals of the three Co atoms, and $p$ orbitals of the two Sn atoms and the two S atoms was constructed using the Wannier90 package \cite{MOSTOFI20142309}. The hybridization between the $d$ orbitals and the $p$ orbitals, together with the spin-orbit coupling effect, is included in our model. Only $3d$ electrons in Co are treated as correlated ones in DFT+DMFT calculations. We chose the fully localized form $\Sigma_{\rm DC}=U(n_{d}^{0}-\frac{1}{2})-\frac{1}{2}J(n_{d}^{0}-1)$, where $n_{d}^{0}$ is nominal occupation of $3d$ orbitals, as the ``double-counting'' scheme.

We used the hybridization expansion version of the continuous-time quantum Monte Carlo (HYB-CTQMC) method implemented in the iQIST package \cite{Gull, Huang:2015cc} as the impurity solver. The local on-site Coulomb interactions are parameterized by the Slater integrals $F^0$, $F^2$ and $F^4$. Hubbard $U$ and Hund's coupling $J$ amount to $U=F^0$, $J=(F^2+F^4)/14$. The constrained DFT calculations suggest $U=5.1$ eV and $J=0.9$ eV for Co$^{2+}$ in CoO \cite{PhysRevB.82.045108}. Besides, the experimental optical absorption data indicate $J_{\rm H}\approx 0.8$ eV \cite{PhysRev.116.281}. Thus, in order to check the effective $U$ and $J$ related to the renormalization factor $\mathcal{Z}$, we fixed the ratio of $J/U=0.2$ to change $U$, which was also used in $d^7$ cobalt compounds study \cite{PhysRevB.97.014407}. We only keep the density-density terms of the Coulomb interactions for computational efficiency. The inverse temperature is $\beta=1/(K_{\rm B}T)$ = 40 eV$^{-1}$. The standard deviation of the self-energy is less than 0.03 in the last self-consistent loop. We used the analytical continuation method introduced by K. Haule \cite{PhysRevB.81.195107} to extract the self-energy $\Sigma({\omega})$ on real axis from the Matsubara self-energy $\Sigma({\mathrm{i}\omega})$ obtained from CTQMC. 
 
In order to study the topological electronic structure of Co$_3$Sn$_2$S$_2$, we calculated the momentum-resolved spectra, which is defined as
\begin{eqnarray} 
A(k,\omega) = -\frac{1}{\pi}\Im{\left[ \frac{1}{ \omega+\mu-H_0(k)-\tilde{\Sigma}(k,\omega) } \right]}
\end{eqnarray}
where, $H_0(k)$ is the non-interaction Hamiltonian at each $k$-point from DFT calculation, $\tilde{\Sigma}(k,\omega)=\hat{P}_{k}(\Sigma(\omega)-\Sigma_{dc})$, $\hat{P}_{k}$ are the projection operators.

The low-energy quasiparticle (QP) behaviour is described by the following QP Hamiltonian,
\begin{eqnarray} 
H_{\rm QP} = H_0 - \mu + \Re{\tilde{\Sigma}(0)}
\end{eqnarray}
Our (001) surface electronic structure, Fermi arcs and Berry curvature were calculated based on the low-energy QP Hamiltonian (See Supplementary Note 5). The surface spectra (i.e., Fermi arcs) were calculated by using the iterative Green's function method \cite{Sancho_1985} as implemented in the WannierTools package \cite{WU2017}. The QP band structure of FM Co$_3$Sn$_2$S$_2$ obtained from the QP Hamiltonian here cannot totally capture the effect of electronic correlations---the reduction of its Drude spectral weight \cite{Kotliar3}.

The real part of the optical conductivity $\sigma_{1}^{\rm QP}$(\emph{$\omega$}) contributed by the direct optical transitions between the calculated quasiparticle bands in the main text was calculated by the Kubo-Greenwood formula as implemented in the Wannier90 package \cite{Basov3, Martin Dressel, MOSTOFI20142309}.

\vspace{1mm} 
\noindent \textbf{Data availability.} Data measured or analyzed during this study are available from the corresponding author on reasonable request.

%


\vspace{2mm} 
\noindent \textbf{Acknowledgements}

\noindent We thank Xi Dai, Hongming Weng, Rui Yu, Quansheng Wu and Xiaoyu Deng for very helpful discussions. The authors acknowledge support from the National Key Research and Development Program of China (Projects No. 2017YFA0304700, No. 2016YFA0300600, No. 2017YFA0302901, No. 2016YFA0300504, No. 2017YFA0303800, No. 2016YFA0302400, and No. 2018YFA0307000), the strategic Priority Research Program of Chinese Academy of Sciences (Project No. XDB33000000), the Pioneer Hundred Talents Program of the Chinese Academy of Sciences, the National Natural Science Foundation of China (Projects No. 11604273, No. 11774399, No. 11574394, No. 11774423, No. 11822412, No. 11721404, and No. 11874022), Longshan Academic Talent Research Supporting Program of SWUST (Project No. 17LZX527) and Beijing Natural Science Foundation (Project No. Z180008). A. S. thanks the support of SNSF NCCR MARVEL and QSIT grants and of Microsoft Research and SNSF Professorship. J. Z. acknowledges the Pauli Center for funding his visit in UZH.

\vspace{2mm} 
\noindent \textbf{Author contributions}

\noindent $^{\dagger}$Y.X., J.Z. and C.Y. contributed equally to this work. \noindent Z.-G.C. conceived and supervised this project. Y.X. and X.H. carried out the optical experiments. J.Z. did first-principle and many-body calculations. C.Y., Q.W., Q.Y., H.L. and Y.S. grew the single crystals. Z.-G.C., J.Z., Y.W., E.L., L.W., G.X., L.L., A.S. and J.L. analyzed the data. Z.-G.C. wrote the paper.  *E-mail address: zgchen@iphy.ac.cn

\vspace{2mm} 
\noindent \textbf{Competing interests}

\noindent The authors declare no competing interests.

\end{document}